# A Divide and Conquer Strategy for Musical Noise-free Speech Enhancement in Adverse Environments

Md Tauhidul Islam, Celia Shahnaz*, *Member, IEEE,* Wei-Ping Zhu, *Senior Member, IEEE,* and M. Omair Ahmad, *Fellow, IEEE*

*Abstract*—A divide and conquer strategy for enhancement of noisy speeches in adverse environments involving lower levels of SNR is presented in this paper, where the total system of speech enhancement is divided into two separate steps. The first step is based on noise compensation on short time magnitude and the second step is based on phase compensation. The magnitude spectrum is compensated based on a modified spectral subtraction method where the cross-terms containing spectra of noise and clean speech are taken into consideration, which are neglected in the traditional spectral subtraction methods. By employing the modified magnitude and unchanged phase, a procedure is formulated to compensate the overestimation or underestimation of noise by phase compensation method based on the probability of speech presence. A modified complex spectrum based on these two steps are obtained to synthesize a musical noise free enhanced speech. Extensive simulations are carried out using the speech files available in the NOIZEUS database in order to evaluate the performance of the proposed method. It is shown in terms of the objective measures, spectrogram analysis and formal subjective listening tests that the proposed method consistently outperforms some of the state-of-the-art methods of speech enhancement for noisy speech corrupted by street or babble noise at very low as well as medium levels of SNR.

## I. Introduction

PRESENCE of background noise degrades the performance of speech communication, speech analysis, and speech recognition systems. For proper operation of such systems in a practical noisy environment, it is desirable to improve the intelligibility and quality of the noisy speech. In order to attain this goal by reducing noise in the noisy speech, various speech enhancement methods, namely, spectral subtraction [1], [2], [3], [4], [5], minimum mean square error (MMSE) estimator [6], [7], subspace based methods [8], [9], wavelet domain based thresholding methods [10], [11], [12], [13], Wiener filtering [14], and Kalman filltering [15] have been reported in the literature. In several of these methods, the analysis-modification-synthesis (AMS) framework [16], [17], [18], [19] is employed for reconstructing the original speech after performing the enhancement operation.

In speech analysis, it is commonly believed that human auditory system is phase-deaf, i.e., it ignores the phase spectrum and considers only the magnitude spectrum. That is why in the conventional spectral subtraction based speech enhancement methods mentioned above, for synthesizing a clean speech, operations are performed only on the short-time magnitude spectrum and an unaltered short-time phase spectrum is maintained. Recently, it has been shown that the phase spectrum is also useful in speech analysis [20], [21], [22].

Among all the methods mentioned above, spectral subtraction has been widely used due to its noise suppression capability with simple computation. In Boll's method [1] of spectral subtraction, the noise spectrum is estimated from the non-speech frames and subtracted from the noisy speech spectrum in the current frame. This simple formulation for enhancing noisy speech comes with prices. If too much noise is subtracted from the noisy speech spectrum, it creates speech distortion. On the other hand, if less noise is subtracted, the enhanced speech remains noisy. For subtracting the proper amount of noise, lots of methods have been proposed such as [23], [24]. Another problem with spectral subtraction is the musical noise, which arises because of raising negative values in the resulting spectrum to zero [25]. Sometimes musical noise is more disturbing than the original noise. To solve the problem of musical noise in [25], the authors proposed to floor the negative spectrum values to some other values than zero.

Spectral subtraction is based on the assumption that the noise and clean speech spectra are totally independent and the cross correlation between them is zero, which is incorrect for most of the practical cases. In [26], the authors show that the cross terms keep crucial impact on the performance of the speech enhancement, when the signal to noise ratio of the noisy speech is less than or near 0 dB. Several attempts have been taken to consider the cross terms for speech enhancement such as [27], [26].

Most of the speech enhancement methods discussed above performs well in high or reasonable SNR levels. But a very few methods have been proposed to cope up with low SNR environments such as [28], [13].

In this paper, we will address the above mentioned problems using a two step formulation. The first step is based on ob-

Md Tauhidul Islam is with the department of Electrical and Computer Engineering, Texas A&M University, College Station, Texas-77843, United States of America, phone: +19797394332, email: tauhid@tamu.edu, and Celia Shahnaz is with the department of Electrical and Electronic Engineering, Bangladesh University of Engineering and Technology, Dhaka-1000, Bangladesh, phone: +8801928568547, email: celia.shahnaz@gmail.com. Wei-Ping Zhu and M. Omair Ahmad are with department of Electrical and Computer Engineering, Concordia University, Montreal, Quebec H3G 1M8, Canada, phone: +1(514)848-2424 ext. 4132,+1(514)848-2424 ext. 3075, email: {weiping,omair}@ece.concordia.ca.

taining a crude estimate of the clean speech spectrum through a modified spectral subtraction method, where we consider the cross-terms between the speech and noise spectrum as non-zero. The second step is based on phase compensation which uses a probabilistic approach to calculate how much compensation should be imposed on the phase spectrum of the noisy speech. An enhanced complex spectrum is obtained by pairing the modified magnitude spectrum from the first step and modified phase spectrum from the second step. Both of the steps produce non-negative results which allow the proposed method to enhance the noisy speech without introducing the musical noise. The proposed method is shown to be effective in producing good results even for noisy speeches with very low SNR levels.

The paper is organized as follows. Section II presents the problem formulation and proposed method. Section III describes the results. Concluding remarks are presented in Section IV.

## II. PROBLEM FORMULATION AND PROPOSED METHOD

In any AMS framework, at first, noisy speech frames are transformed by a transformation method. Then modifications are carried out in the transformed domain and finally, the inverse transform of the transformation method followed by the overlap-add method is performed to reconstruct the enhanced speech. The proposed method is based on the AMS framework where speech is analyzed, modified and synthesized frame wise.

In the presence of additive noise d[n], a clean speech signal x[n] gets contaminated and produces noisy speech y[n]. The noisy speech can be segmented into overlapping frames by using a sliding window. $l^{th}$ windowed noisy speech frame can be expressed in the time domain as

$$y^l[n] = x^l[n] + d^l[n], l = 1, \ldots, L, \quad (1)$$

where $L$ is the total number of speech frames. If $Y^l[k]$, $X^l[k]$ and $D^l[k]$ are the short-time Fourier transform (STFT) representations of $y^l[n]$, $x^l[n]$ and $d^l[n]$, respectively, we can write

$$Y^l[k] = X^l[k] + D^l[k], \quad (2)$$

where $k = 0, 1, 2, \ldots N - 1$, $N$ is the length of a frame in samples. The $N$-point FFT, $Y^l[k]$ of $y^l[n]$ can be computed as

$$Y^l[k] = \sum_{n=0}^{N-1} y^l[n] e^{-\frac{j2\pi nk}{N}}. \quad (3)$$

The Fourier transform of the noisy speech frame, $Y^l[k]$ is modified in the proposed method to obtain an estimate of the clean speech spectrum.

An overview of the proposed speech enhancement method is shown by a block diagram in Fig. 1. It is seen from Fig. 1 that short-time Fourier transform (STFT) is first applied to each input speech frame. The magnitude of the Fourier spectrum is compensated in a modified spectral subtraction method, which we call M-step. The modified magnitude from M-step is then combined with unchanged phase to obtain the modified complex spectrum. Using inverse fast Fourier transform (IFFT) and overlap and add, an intermediate speech signal is obtained. The spectrum of the intermediate speech is sent to P-step, which consists of phase spectrum compensation (PSC) [20]. PSC modifies the phase spectrum based on the probability of speech presence in the intermediate speech. Using the modified phase spectrum with the modified magnitude spectrum from the first step, we obtain an enhanced complex spectrum. Finally, using IFFT and overlap and add, an enhanced speech is constructed. The full AMS process is done for both steps to get full flexibilities of using different window sizes and parameters.

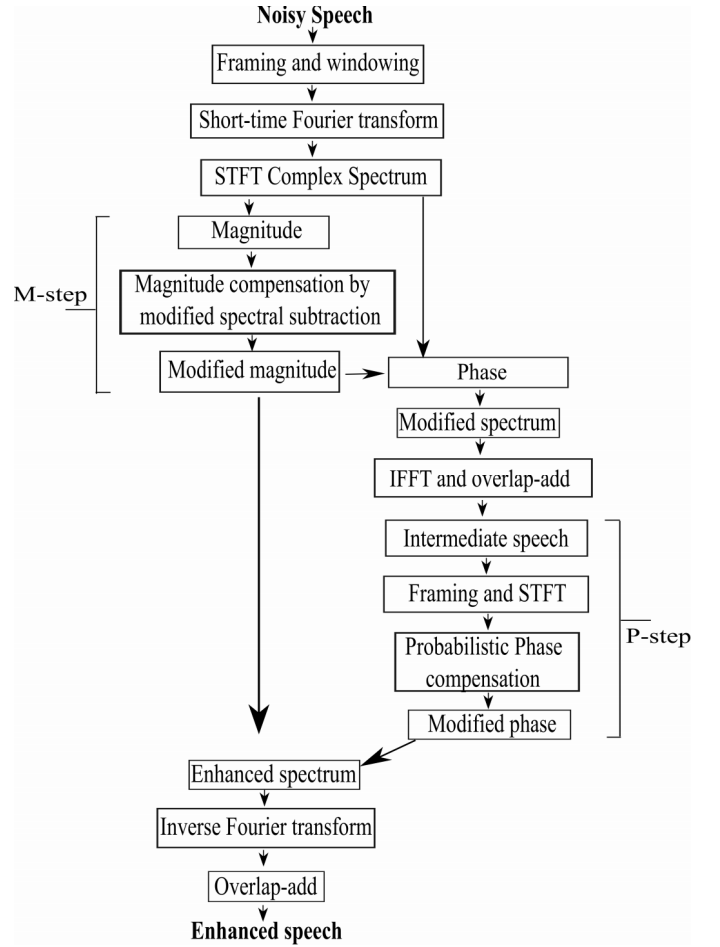

Fig. 1: Block diagram of the proposed method.

### A. M Step: magnitude compensation based on modified spectral subtraction

To get the power spectrum of the $l^{th}$ frame of noisy speech, we multiply $Y^l[k]$ in the (2) by its conjugate $(Y^l[k])^*$. In doing so, (2) becomes

$$Y^l[k]^2 = |X^l[k]|^2 + |D^l[k]|^2 + X^l[k] \cdot (D^l[k])^* + (X^l[k])^* \cdot D^l[k], \quad (4)$$

which can be written as

$$\begin{aligned}|X^l[k]|^2 =& |Y^l[k]|^2 - |D^l[k]|^2 - X^l[k] \cdot (D^l[k])^* - (X^l[k])^* \cdot D^l[k] \\
=& ||Y^l[k]|^2 - |D^l[k]|^2 - (Y^l[k] - D^l[k]) \cdot (D^l[k])^* \\
& - (Y^l[k] - D^l[k])^* \cdot D^l[k]|. \quad (5)\end{aligned}$$

In this equation, we see that even if $D^l[k]$ is greater than $Y^l[k]$, $|X^l[k]|^2$ does not become negative. The other two negative term becomes positive and makes $|X^l[k]|^2$ positive. We can now define a gain function for the modified spectral subtraction method in (5) as

$$|Z^l[k]|^2 = H_{MSS}^l[k] Y^l[k]^2, \qquad (6)$$

where $Z^l[k]$ is an estimate of $X^l[k]$ from M-step and the gain function for the modified spectral subtraction method is denoted as $H_{MSS}^l$. In (6), $H_{MSS}^l[k]$ is defined as

$$H_{MSS}^l[k] = \sqrt{\left|1 - \frac{|D^l[k]|^2}{|Y^l[k]|^2} - \chi\right|}. \qquad (7)$$

In (7), $1 - \frac{|D^l[k]|^2}{|Y^l[k]|^2}$ in right hand side is the square of spectral gain of conventional spectral subtraction method and $\chi$-term is the cross correlation between $X^l[k]$ and $D^l[k]$. We can simplify (7) into

$$H_{MSS}^l[k] = \sqrt{|(H_{SS}^l)^2 - \chi|}, \qquad (8)$$

where the gain function for the classical spectral subtraction method is denoted as $H_{SS}^l$ and $\chi$ is defined as

$$\chi = \frac{(Y^l[k] - D^l[k]) \cdot (D^l[k])^* + (Y^l[k] - D^l[k])^* \cdot D^l[k]}{|Y^l[k]|^2}. \qquad (9)$$

Please note that a voice activity detector is used in the proposed scheme from [1] for detecting the speech and silence frames. We obtain an modified complex spectrum by aggregating the modified magnitude with the unchanged phase of the noisy speech spectrum.

$$Z^l[k] = |Z^l[k]| e^{j \angle Y^l[k]}. \qquad (10)$$

A noisy speech file corrupted by 10 dB babble noise from NOIZEUS database is processed for classical spectral subtraction method [1] and modified spectral subtraction method to show their qualitative difference. The spectrograms of the processed files and the clean speech are shown in Fig. 2. From this figure, we see that classical spectral subtraction produces a cleaner result but removes some speech harmonics. On the other hand in the modified spectral subtraction method, the speech harmonics are preserved, although there are some residual noises. The residual noise is taken care of by the next P-step. The P-step is based on phase compensation where we use a probabilistic approach to modify the result from the M-step. The next P-step is expected to modify the output from M-step in two ways.

1) If the noise estimation process gives an underestimation of true noise, the next step should reduce the output to compensate for the true noise.
2) If the estimated noise gives overestimation of noise, the next step should increase the output for compensating it.

After using IFFT on $Z^l[k]$ and overlap and add of real part of the resulting signal, we obtain time-domain intermediate speech $z[n]$.

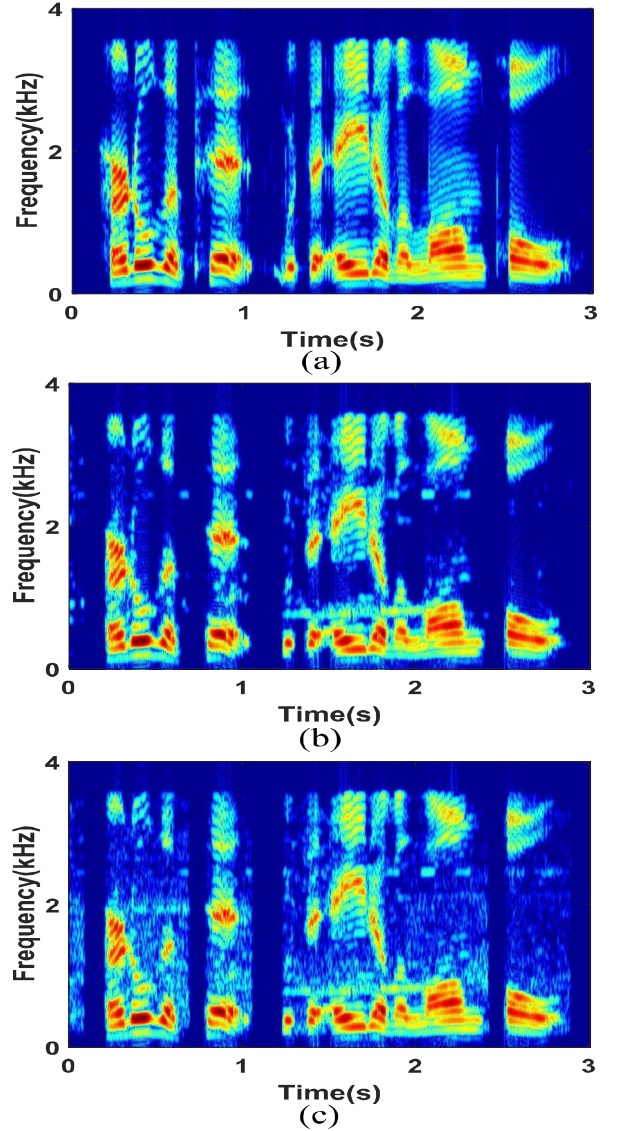

Fig. 2: Comparison between the modified spectral subtraction and classical spectral subtraction. Spectrograms of (a) clean speech; Spectrograms of noisy speeches processed by (b) classical spectral subtraction (c) modified spectral subtraction.

*B. P Step: phase compensation based on probability of speech presence*

If we apply STFT on $z[n]$, we obtain $Z^\tau[k]$, where $\tau$ is the frame number for P-step. In P-step, the modified complex spectrum $Z^\tau[k]$ is modified in such a way that the low energy component cancel out more than the high energy components. The modified complex spectrum thus obtained is a better representation of $X^\tau[k]$.

$$\widehat{X}^\tau[k] = |Z^\tau[k]| e^{j \angle (Z^\tau[k] + \phi^\tau[k])}. \qquad (11)$$

$z^\tau[n]$, $\tau^{th}$ frame of the intermediate speech, is a real valued signal and therefore, its FFT is conjugate symmetric, i.e.,

$$Z^\tau[k] = \{Z^\tau[N_\tau - k]\}^*, \qquad (12)$$

where $N_\tau$ is the number of samples in a frame in P-step. The conjugates can be obtained as a result of applying FFT on $z^\tau[n]$. The conjugates arise naturally from the symmetry of the magnitude spectrum and anti-symmetry of the phase spectrum. During IFFT operation as needed for synthesis of enhanced speech, the conjugates are summed together to produce larger real valued signal. If the conjugates are modified, the degree to which they sum together can be influenced and this can be contributed constructively or destructively to the reconstruction of the enhanced time domain speech. For this purpose, we formulate a phase spectrum compensation function as given by

$$\phi^\tau[k] = \mu \rho^\tau[k] \Lambda[k] V^\tau[k], \quad (13)$$

where $\mu$ is a constant determined empirically, $V^\tau$ is the estimate of noise spectrum in $\tau^{th}$ frame determined as the root mean square of $Z^\tau$, where $Z^\tau = (Z^\tau[1], \ldots Z^\tau[N_\tau])^T$ [20], $\rho^\tau[k]$ is a real-valued constant that is expected to be dependent on the magnitudes of clean speech and noise spectra. Instead of taking it as a constant like [20] or SNR-dependent as [22], we take it as dependent on the probability of speech presence in $\tau^{th}$ frame [5] as

$$\rho^\tau[k] = \sqrt{1 - P_{local}^\tau[k] P_{global}^\tau[k] P_{frame}(\tau)}, \quad (14)$$

where $P_{local}^\tau[k]$ and $P_{global}^\tau[k]$ are the probability of speech presence in a local and global window determined from the following equation,

$$P_\psi^\tau[k] = \begin{cases} 0, & \text{if } \xi_\psi^\tau[k] \leq \xi_{min}, \\ 1, & \xi_\psi^\tau[k] \geq \xi_{max}, \\ \frac{\log(\xi_\psi^\tau[k]/\xi_{min})}{\log(\xi_{max}/\xi_{min})}, & \text{otherwise,} \end{cases} \quad (15)$$

where the subscript $\psi$ denotes either "local" or "global" and $\xi_\psi^\tau[k]$ represents either "local" or "global" mean values of the *apriori* SNR [28], [6]. $\xi_\psi^\tau[k]$ is defined as

$$\xi_\psi^\tau[k] = \sum_{i=-W_\psi}^{i=W_\psi} h_\psi(i) \xi^\tau[k-i], \quad (16)$$

where $h_\psi$ is a window function of length of $2W_\psi + 1$. In this equation, $\xi^\tau[k]$ is defined as

$$\xi^\tau[k] = (1 - \alpha_\xi) \hat{\gamma}^\tau[k], \quad (17)$$

where $\alpha_\xi$ is a constant and $\hat{\gamma}^\tau[k]$ is the estimated *a posteriori* SNR.

In (14), the probability of speech presence in $\tau^{th}$ frame, $P_{frame}(\tau)$ is determined as

$$P_{frame}(\tau) = \begin{cases} 0, & \text{if } \xi_{frame}(\tau) < \xi_{min}, \\ 1, & \text{if } \xi_{frame}(\tau) > \xi_{frame}(\tau-1) \text{ and } \\ & \xi_{frame}(\tau) > \xi_{min}, \\ \mu(\tau), & \text{otherwise,} \end{cases} \quad (18)$$

where $\mu(\tau)$ is determined by

$$\mu(\tau) = \begin{cases} 0, & \text{if } \xi_{frame}(\tau) \leq \xi_{peak} \xi_{min}, \\ 1, & \text{if } \xi_{frame}(\tau) \geq \xi_{peak} \xi_{max}, \\ \frac{\log(\xi_{frame}(\tau)/(\xi_{peak} \xi_{min}))}{\log(\xi_{max}/\xi_{min})}, & \text{otherwise.} \end{cases} \quad (19)$$

In (15), (18) and (19), $\xi_{min}$, $\xi_{max}$ and $\xi_{peak}$ are empirically determined constants and $\xi_{frame}(\tau)$ is defined as

$$\xi_{frame}(\tau) = \frac{1}{N_\tau} \sum_{k=0}^{N_\tau - 1} \xi^\tau[k]. \quad (20)$$

To understand the effect of this probabilistic approach of determination of phase compensation function, we plot the noise to signal ratio and speech presence probability in Fig. 3 for a randomly chosen frame of a noisy speech. From this figure, we see that at the sample points, where the noise to signal ratio is low, the probability of speech presence is high, which ensures that less phase compensation is imposed on the noisy speech spectrum so that speech harmonics does not get distorted. On the other hand, at the sample points, where the noise to signal ratio is high, the probability of speech presence is low. This situation implies that there are severe noises at these sample points. A large phase compensation in the proposed method ensures that the noise is removed completely.

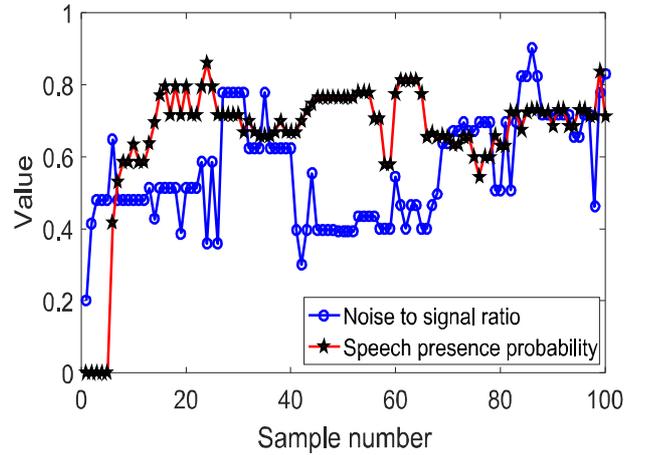

Fig. 3: Noise to signal ratio and speech presence probability in a randomly chosen frame.

In (13), $\Lambda[k]$ is defined as

$$\Lambda[k] = \begin{cases} 1, & \text{if } 0 < \frac{k}{N} < \frac{1}{2}, \\ -1, & \text{if } \frac{1}{2} < \frac{k}{N} < 1, \\ 0, & \text{otherwise.} \end{cases} \quad (21)$$

From this equation, we realize that zero weighting is assigned to the values of k corresponding to the non-conjugate vectors of FFT, such as at $k = 0$ and at $k = N/2$ if $N$ even. Since the estimate of noise magnitude spectrum is symmetric, introduction of the weighting function defined by (21) produces an anti-symmetric compensation function in (13) that acts as the cause for changing the angular phase relationship in order to achieve noise cancellation in synthesis step in the proposed phase compensation scheme.

As we discussed in the previous section, the incorrect noise estimation can give rise to two types of error, i.e., underestimation of noise can keep the signal vector high and overestimation of noise can decrease the signal vector more than necessary. We show these two situations with help of a vector diagram in Fig. 4, where both the time and frequency indices are omitted of the vectors for convenience and clarity. As the phase compensation function $\phi$ is a scaled noise vector, it has the same phase as noise vector. From Fig. 4, we see that the conjugate components of additive phase compensation function are denoted as $\phi_+$ and $\phi_+^*$ and the conjugate components of subtractive phase compensation function are denoted as $\phi_-$ and $\phi_-^*$. If $Z^\tau[k]$ comes from a underestimation of noise that means still there are some noise in it, noise and phase compensation function $\phi^\tau[k]$ can be thought as additive. On the other hand, if $Z^\tau[k]$ comes from an overestimation of noise, which means that true signal is larger than $Z^\tau[k]$, noise and $\phi^\tau[k]$ can be thought as subtractive. For $\phi^\tau[k]$ to be additive, it should have a phase of $-90^o$ to $90^o$ and to be subtractive, it should have phase of $90^o$ to $270^o$. Now we will discuss in next paragraph how phase compensation works for these two cases.

Explanation for two cases of single conjugate pair and their corresponding modifications, i.e., when the estimated speech vector from first step is greater and smaller than the phase compensation function are presented in Figs. 5 and 6 for additive and subtractive phase compensation functions, where both the time and frequency indices are again omitted of the vectors for convenience and clarity. We will denote the two conjugates of $Z^\tau[k]$ as $\overrightarrow{Z}$ and $\overrightarrow{Z^*}$, of additive phase compensation function as $\overrightarrow{\phi_+}$ and $\overrightarrow{\phi_+^*}$, of subtractive phase compensation function as $\overrightarrow{\phi_-}$ and $\overrightarrow{\phi_-^*}$ and of $\widehat{X}^\tau[k]$ as $\overrightarrow{X}$ and $\overrightarrow{X^*}$.

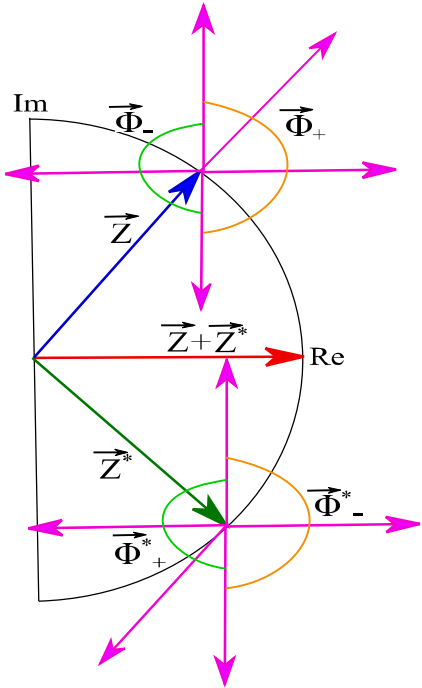

Fig. 4: Vector diagram showing the phase compensation functions for underestimated and overestimated noises with the intermediate speech spectrum.

*1) Case 1: when phase compensation function is additive:* In Fig. 5, the first row (a) shows the first option when $|\overrightarrow{Z}|$ is greater than $|\overrightarrow{\phi_+}|$. First column shows $\overrightarrow{Z}$, its conjugate and the sum vector of $\overrightarrow{Z}$ and its conjugate for both of the cases. The second Column shows $\overrightarrow{\phi_+}$, its conjugate with $\overrightarrow{Z}$. For advantage of analysis, we assume that $\overrightarrow{\phi_+}$ has an angle of $0^o$. For the case where of $\overrightarrow{\phi_+}$ has any other angle between $-90^o$ to $90^o$, the analysis does not change. According to (13), $\overrightarrow{\phi_+}$ is added with $\overrightarrow{Z}$ which is shown in second column of (a) in Fig. 5. The conjugate of $\overrightarrow{\phi_+}$ has an opposite direction to $\overrightarrow{\phi_+}$. After vector addition of $\overrightarrow{Z}$ and $\overrightarrow{\phi_+}$ and $\overrightarrow{Z^*}$ and $\overrightarrow{\phi_+^*}$, we obtain $\overrightarrow{X}$ and $\overrightarrow{X^*}$ shown in third column of (a). In the forth column of (a) of Fig. 5, we show the resulting clean signal vector $\overrightarrow{X} + \overrightarrow{X^*}$ and its magnitude. The case where $|\overrightarrow{\phi_+}|$ is larger than $|\overrightarrow{Z}|$ is shown in (b) of the same figure. The same vector addition process is followed in (b) as (a). We see from forth column of (b) that the resulting clean signal vector obtained from this case is significantly less than the case when $|\overrightarrow{Z}|$ is greater than $|\overrightarrow{\phi_+}|$. These results are compliant with [20].

*2) Case 2: when phase compensation function is subtractive:* In Fig. 6, the first row (a) shows the case when $|\overrightarrow{Z}|$ is greater than $|\overrightarrow{\phi_-}|$ and second row shows the case when $|\overrightarrow{Z}|$ is smaller than $|\overrightarrow{\phi_-}|$. First column shows $\overrightarrow{Z}$, its conjugate and the sum vector of $\overrightarrow{Z}$ and its conjugate for both of the cases. The second Column shows the addition of $\overrightarrow{Z}$ with $\overrightarrow{\phi_-}$ and $\overrightarrow{Z^*}$ with $\overrightarrow{\phi_-^*}$. For advantage of analysis, in this case, we assume that $\overrightarrow{\phi_-}$ has an angle of $180^\circ$, which is shown in second column of Fig. 6. The third column is used to show the vector summation of $\overrightarrow{Z}$ and $\overrightarrow{\phi_-}$. In the forth column of (a), the obtained clean signal vector $\overrightarrow{X} + \overrightarrow{X^*}$ is shown, whose magnitude is larger than the magnitude of $\overrightarrow{Z}$. This is how the proposed phase compensation function resolves the issue of overestimation of noise. The case where $|\overrightarrow{\phi_-}|$ is larger than $|\overrightarrow{Z}|$ is shown in second row (b) of the same figure. For this case, we obtain the resultant clean signal vector $\overrightarrow{X} + \overrightarrow{X^*}$ very small. It is rational, since most of the time, obtaining very large subtractive noise is because of the inaccurate noise estimation and the intermediate speech spectrum should be highly compensated for that.

We realize from the above discussion and figures that the resulting spectrum obtained from the P-step, $\widehat{X}^\tau[k]$ is produced considering the noise characteristics in the intermediate speech. We also realize from Figs. 5 and 6 that the result obtained from the P-step is always non-negative. Based on this, we can expect the proposed method to enhance the noisy speech without introducing any musical noise.

## C. Resynthesis of enhanced signal

The enhanced speech frame is synthesized by performing the IFFT on the resulting $\widehat{X}^\tau[k]$,

$$\widehat{x}^\tau[k] = Re\left(IFFT\{\widehat{X}^\tau[k]\}\right), \qquad (22)$$

where $Re(\cdot)$ denotes the real part of the number inside it and $\widehat{x}[n]$ represents the the enhanced speech frame. The final enhanced speech signal is synthesized by using the standard overlap and add method [29].

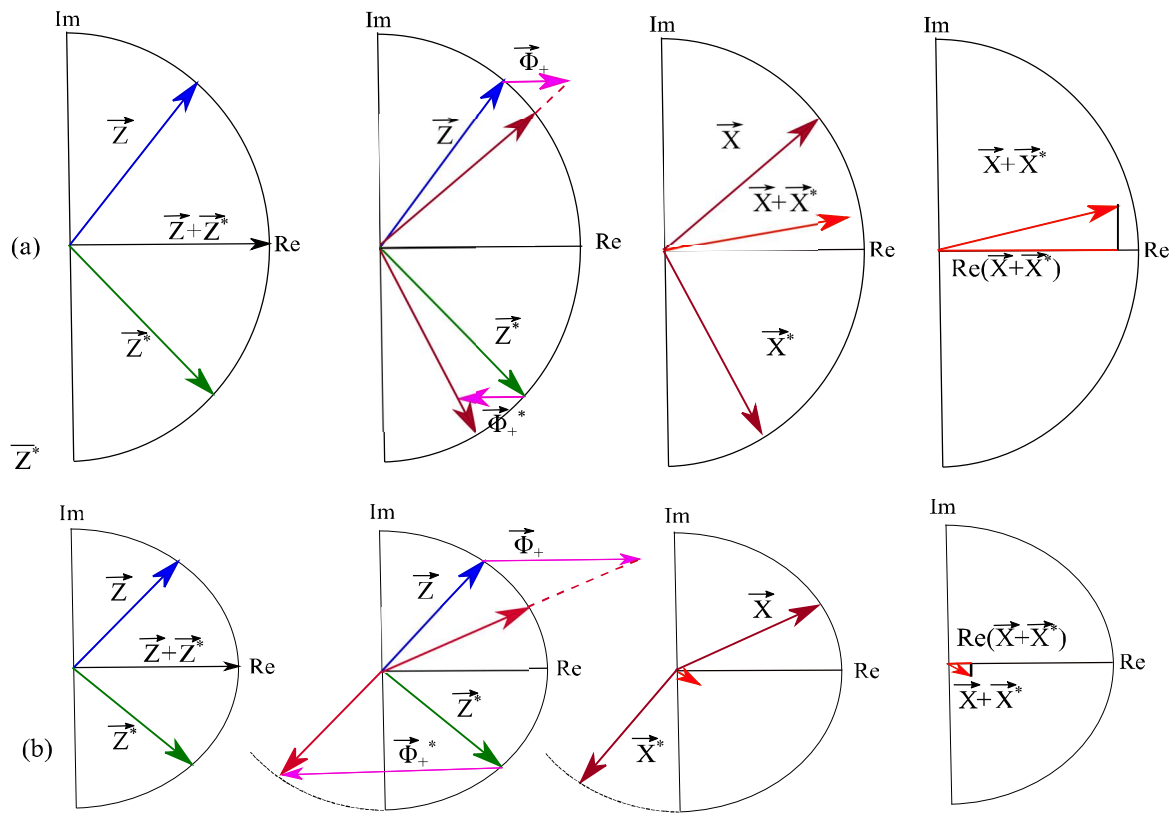

Fig. 5: Phase compensation for additive phase compensation, $\vec{\phi}_+$ (a) when $|\vec{Z}| > |\vec{\phi}_+|$ (b) when $|\vec{Z}| < |\vec{\phi}_+|$.

## III. Results

In this section, a number of simulations is carried out to evaluate the performance of the proposed method.

### A. Implementation

The above proposed method which we call modified spectral subtraction with probabilistic phase compensation (MSPP), is implemented in MATLAB R2016b graphical user interface development environment (GUIDE). The MATLAB software with its user manual is attached as supplementary material with the paper. This software also includes implementation of some classical as well as recent methods, i.e., spectral subtraction (SS) [1], multi-band Spectral Subtraction (MBSS) [24], phase spectrum compensation (PSC) [20] and soft mask estimator with posteriori SNR uncertainty (SMPO) [30]. The implementations of these methods have been taken from publicly available and trusted sources. SS code is taken from https://www.mathworks.com/matlabcentral/fileexchange/7675-boll-spectral-subtraction, MBSS code is taken from https://www.mathworks.com/matlabcentral/fileexchange/7674-multi-band-spectral-subtraction, PSC implementation code is acquired from https://www.mathworks.com/matlabcentral/fileexchange/30815-phase-spectrum-compensation and SMPO code is taken from http://ecs.utdallas.edu/loizou/cimplants/. The MATLAB implementations of the calculation of segmental SNR (SNRSeg) improvement and overall SNR improvement are taken from http://ecs.utdallas.edu/loizou/cimplants/ [31].

TABLE I: Constants used to determine the phase compensation function in P-step

| Constants | Values of constants |
|---|---|
| $\mu$ | 0.6 |
| $\xi_{min}$ | -10 dB |
| $\xi_{max}$ | -5 dB |
| $\xi_{peak}$ | 10 dB |
| $W_{local}$ | 1 |
| $W_{global}$ | 15 |
| $\alpha_\xi$ | 0.7 |

### B. Simulation Conditions

Real speech sentences from the NOIZEUS database are employed for the experiments, where the speech data is sampled at 8 kHz. To imitate a noisy environment, noise sequence is added to the clean speech samples at different signal to noise ratio (SNR) levels ranging from 10 dB to $-30$ dB. Two different types of noises, namely babble and street are adopted from the NOIZEUS database.

In order to obtain overlapping analysis frames in M-step, Hamming windowing operation is performed, where the size of each of the frame is 100 samples with 50% overlap between successive frames. In P-step, Griffin and Lim's modified Hanning window is used and the size of each frame is 256 samples with 25% overlap. The constant, $\beta$ in M-step is taken as 0.7 and the values of used constants to determine the phase compensation function in P-step are given in Table I.

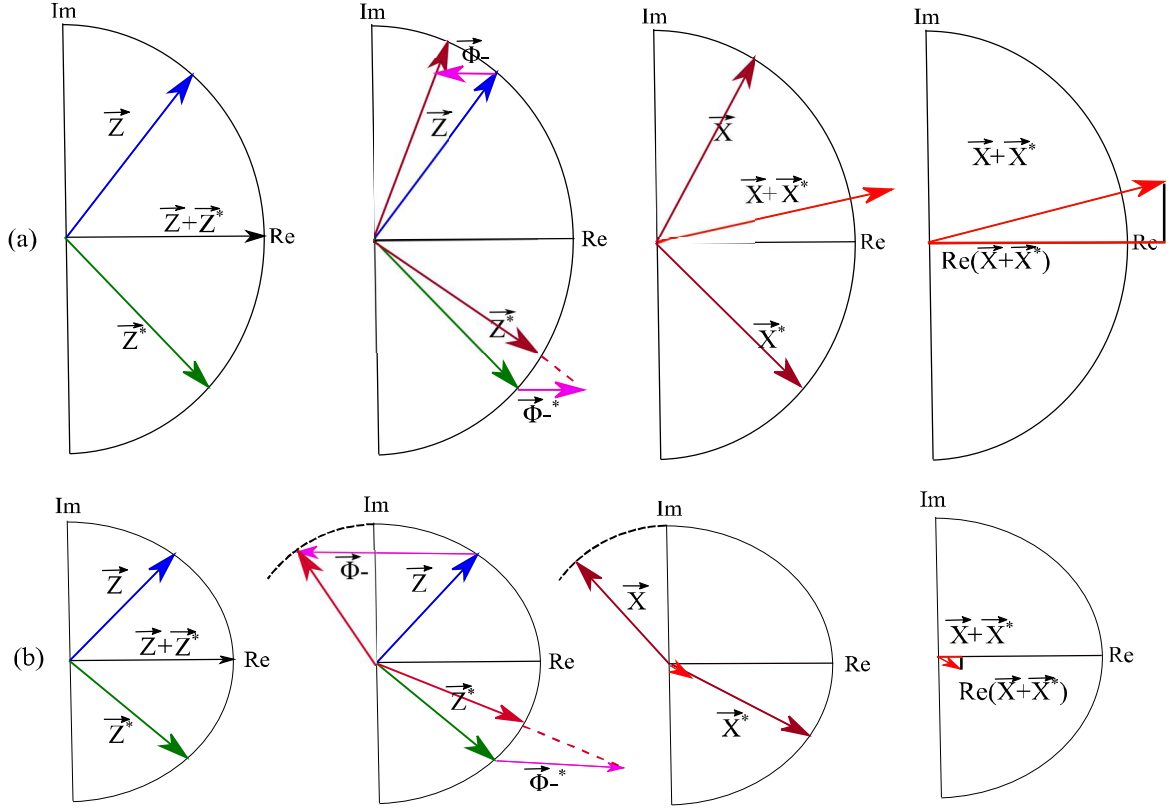

Fig. 6: Phase compensation for subtractive phase compensation, $\vec{\phi_-}$ (a) when $|\vec{Z}| > |\vec{\phi_-}|$ (b) when $|\vec{Z}| < |\vec{\phi_-}|$.

## C. Comparison Metrics

Standard Objective metrics [32], namely, SNRSeg improvement in dB, overall SNR improvement in dB and perceptual evaluation of speech quality (PESQ) are used for the evaluation of the proposed method. The proposed method is subjectively evaluated in terms of the spectrogram representations of the clean speech, noisy speech and enhanced speech. Formal listening tests are also carried out in order to find the analogy between the objective metrics and subjective sound quality. The performance of the proposed MSPP method is compared with MBSS [24], PSC [33] and SMPO [30] in both objective and subjective senses.

## D. Objective Evaluation

*1) Results for speech signals with street noise:* SNRSeg improvement, overall SNR improvement and PESQ scores for speech signals corrupted with street noise for MBSS, PSC, SMPO and MSPP are shown in Fig. 7, 8 and Table II.

In Fig. 7, we compare the performance of the proposed MSPP method with MBSS, PSC and SMPO in terms of SNRSeg improvement for the SNR range of $-30$ to $10$ dB. From this figure, we see that for extremely low SNR $-30$ dB, the proposed method provides an SNRSeg improvement of $8$ dB, whereas SMPO, MBSS and PSC provides $1.5$, $1.2$ and $0.5$ dB only. For high SNR as $10$ dB, we see that the SNRSeg improvement for proposed method is low as $1.5$ dB. For other methods, at $10$ dB, the SNRSeg improvement is as low as $0.1$ dB for MBSS. With decrement of SNR, the SNRSeg improvement increases for every methods upto around $-5$ dB. After that we see that proposed method still continues to increase the SNRSeg improvement whereas for other methods, SNRSeg improvement becomes almost constant. It proves that the proposed method performs significantly better than all the other methods in lower SNR cases.

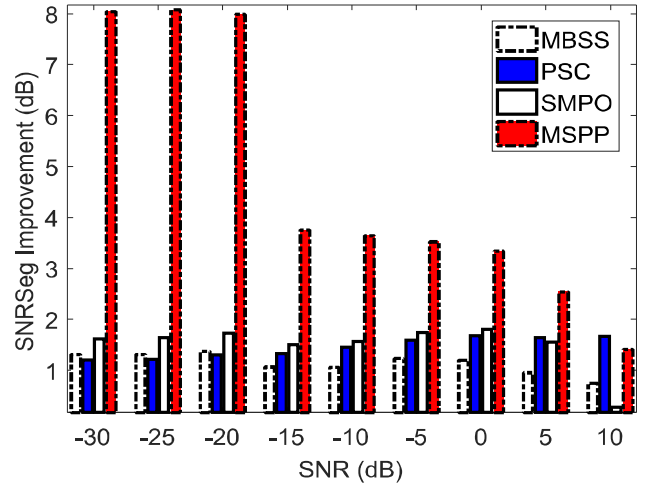

Fig. 7: SNRSeg improvement for different methods in street noise.

Fig. 8 shows the overall SNR improvements in dB as a function of SNR for MSPP and those for the other methods. As shown in the figure, the overall SNR improvements resulting from MBSS and PSC are comparable and relatively smaller for all the SNR levels, whereas MSPP provides better values for most of the SNR levels in comparison to all other methods, which proves the efficacy of MSPP in producing speeches with better quality.

TABLE II: PESQ for different methods in street

| SNR(dB) | MBSS | PSC  | SMPO | MSPP |
|---------|------|------|------|------|
| -30     | 0.05 | 0.09 | 0.35 | 0.30 |
| -25     | 0.47 | 0.62 | 0.97 | 0.81 |
| -20     | 0.56 | 0.78 | 1.01 | 1.05 |
| -15     | 1.15 | 1.42 | 1.45 | 1.30 |
| -10     | 1.37 | 1.56 | 1.57 | 1.51 |
| -5      | 1.51 | 1.59 | 1.51 | 1.65 |
| 0       | 1.69 | 1.72 | 1.69 | 1.83 |
| 5       | 2.07 | 2.17 | 2.57 | 2.54 |
| 10      | 2.38 | 2.56 | 2.68 | 2.65 |

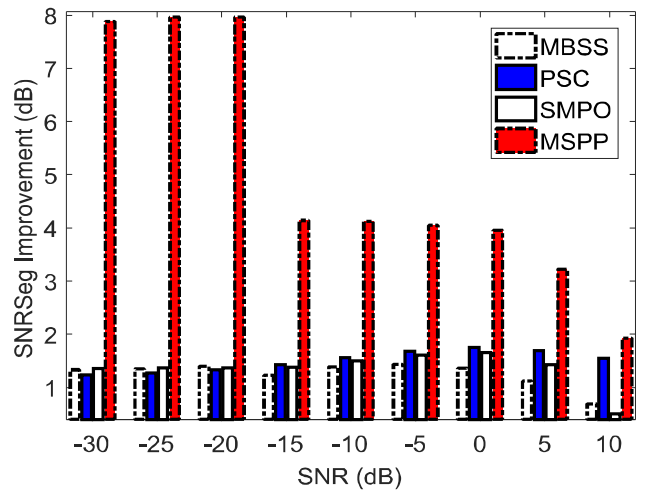

Fig. 9: SNRSeg improvement for different methods in babble noise.

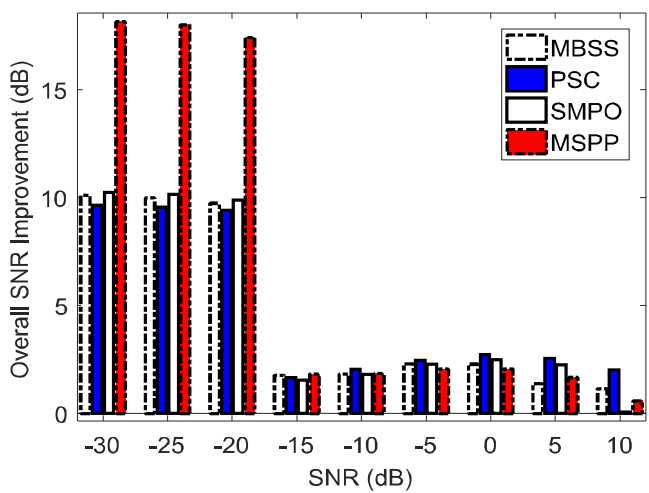

Fig. 8: Overall SNR improvement for different methods in street noise.

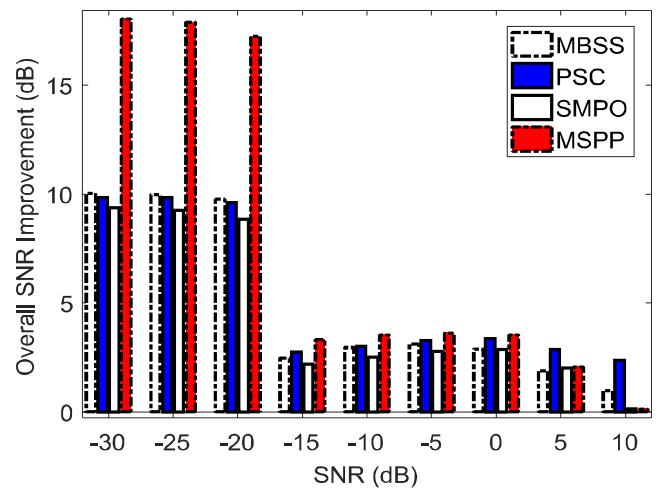

Fig. 10: Overall SNR improvement for different methods in babble noise.

PESQ values for different methods for street noise-corrupted speeches are shown in Table. II. For higher noise as 10 dB, we see that all the methods provide better PESQ. But with the decrement of SNR, PESQ values for all the cases start to fall. The proposed method provides very competitive PESQ values for all SNR levels in comparison to SMPO but performs better than other two competing methods. As PESQ values indicate the perceptual quality of the enhanced speech, this table proves that the proposed method provides better enhanced speeches for street noise corrupted speeches at high as well as low SNRs than MBSS and PSC.

*2) Results for speech signals with multi-talker babble noise:* SNRSeg improvement, overall SNR improvement and PESQ scores for speech signals corrupted with babble noise for MBSS, PSC, SMPO and MSPP are shown in Figs. 9, 10 and 11.

In Fig. 9, we compare the performance of the proposed method in terms of SNRSeg improvement in dB with those of the other methods at different levels of SNR. From this figure, we realize that for highly noisy situation, i.e., $-30$ dB, the proposed method provides a SNRSeg improvement of $5.9$ dB, which is significantly better than other competing methods. The proposed method continues to perform better than other methods in other low levels of SNR as well as higher SNR as 10 dB.

We plot the overall SNR improvement scores for the proposed method and those of the competing methods in Fig. 10 for babble noise-corrupted speeches. From this figure, we see that the MSPP provides better overall SNR improvements for SNR range of $-30$ dB to $0$ dB and competitive improvements in other SNR levels in comparison to other methods.

Mean PESQ values with standard deviations for all the methods for all the thirty files of NOIZEUS database are shown in Fig. 11. From this figure, we see that although the proposed method provides competitive values in comparison to other methods at lower SNRs, it provides better PESQ values

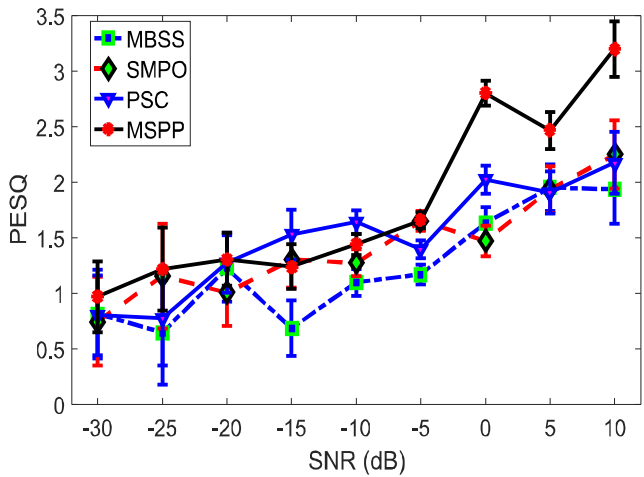

Fig. 11: PESQ for different methods in babble noise.

for high SNRs with smaller standard deviations.

*E. Subjective Evaluation*

To evaluate the performance of the proposed method and other competing methods subjectively, we use two commonly used tools. The first one is the plot of the spectrograms of the output for all the methods and compare their performances in terms of preservation of harmonics and capability to remove noise.

The spectrograms of the clean speech, the noisy speech, and the enhanced speech signals obtained by using the proposed MSPP method and all other methods are presented in Fig. 12 for babble noise-corrupted speech at an SNR of 10 dB. It is obvious from the spectrograms that the proposed method preserves the harmonics significantly better than all other competing methods. The noise is also reduced at every time point for the proposed method which attest our claim of better performance in terms of higher SNRseg improvement, higher overall SNR improvement and higher PESQ values in objective evaluation.

The second tool we used for subjective evaluation of the proposed method and the competing methods is the formal listening test. We add street and babble noises to all the thirty speech sentences of NOIZEUS database at $-15$ to $10$ SNR levels and process them with all the competing methods. We allow ten listeners to listen to these enhanced speeches from these methods and evaluate them subjectively. Following [13] and [34], We use SIG, BAK and OVL scales on a range of $1$ to $5$. The detail of these scales and procedure of this listening test is discussed in [13]. More details on this testing methodology of listening test can be found in [35].

We show the mean scores of SIG, BAK, and OVRL scales for all the methods for speech signals corrupted with street noise in Tables III, IV, and V and for speech signals corrupted with babble noise in Tables VI, VII, and VIII. The higher values of these scores for the proposed method for most of the cases in comparison to other methods clearly attest that the proposed method is better than them in terms of lower signal distortion (higher SIG scores), efficient noise removal (higher BAK scores) and overall sound quality (higher OVL scores) for all SNR levels.

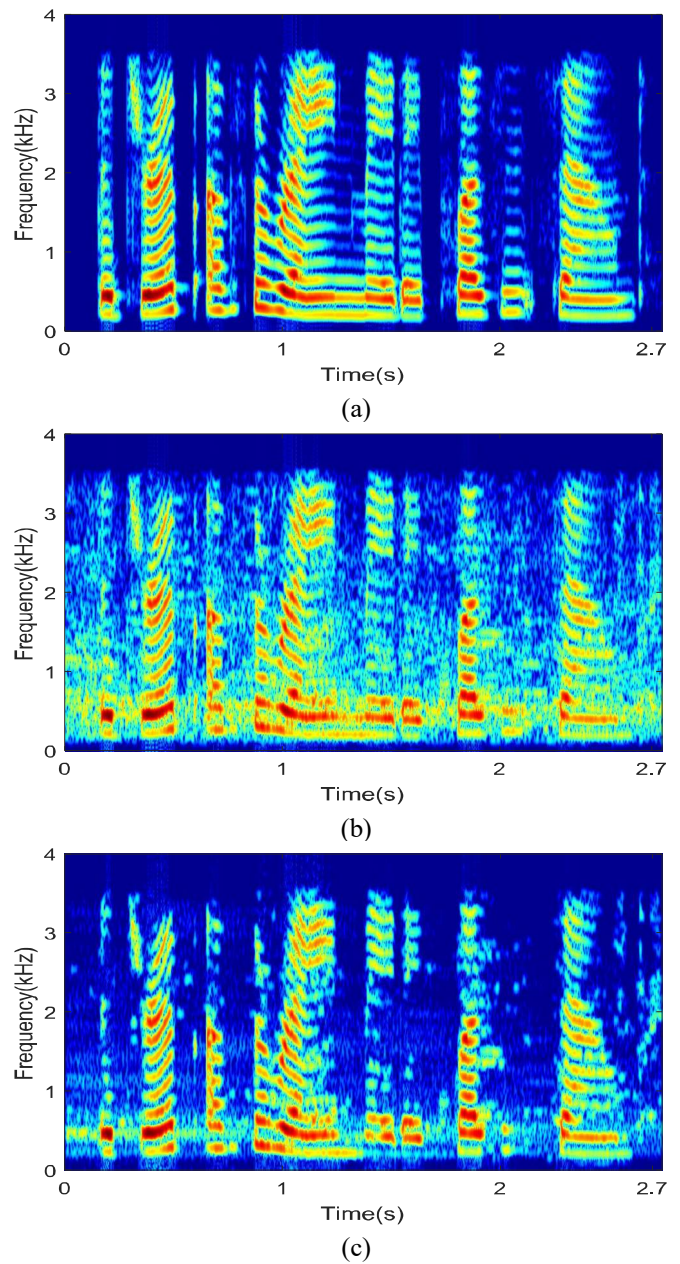

Fig. 12: Spectrograms of (a) clean signal (b) noisy signal with 10 dB babble noise; spectrograms of enhanced speech from (c) MBSS (d) PSC (e) SMPO (f) MSPP.

## IV. CONCLUSIONS

An improved speech enhancement method for enhancing the noisy speech with medium and low SNRs has been proposed in this paper. The proposed method utilizes a modified spectral subtraction and a phase compensation method based on speech presence probability to obtain a better musical noise-free enhanced speech. The proposed method considers the cross-terms between the clean speech and noise for enhancement of the noisy speech. This method is shown to improve the

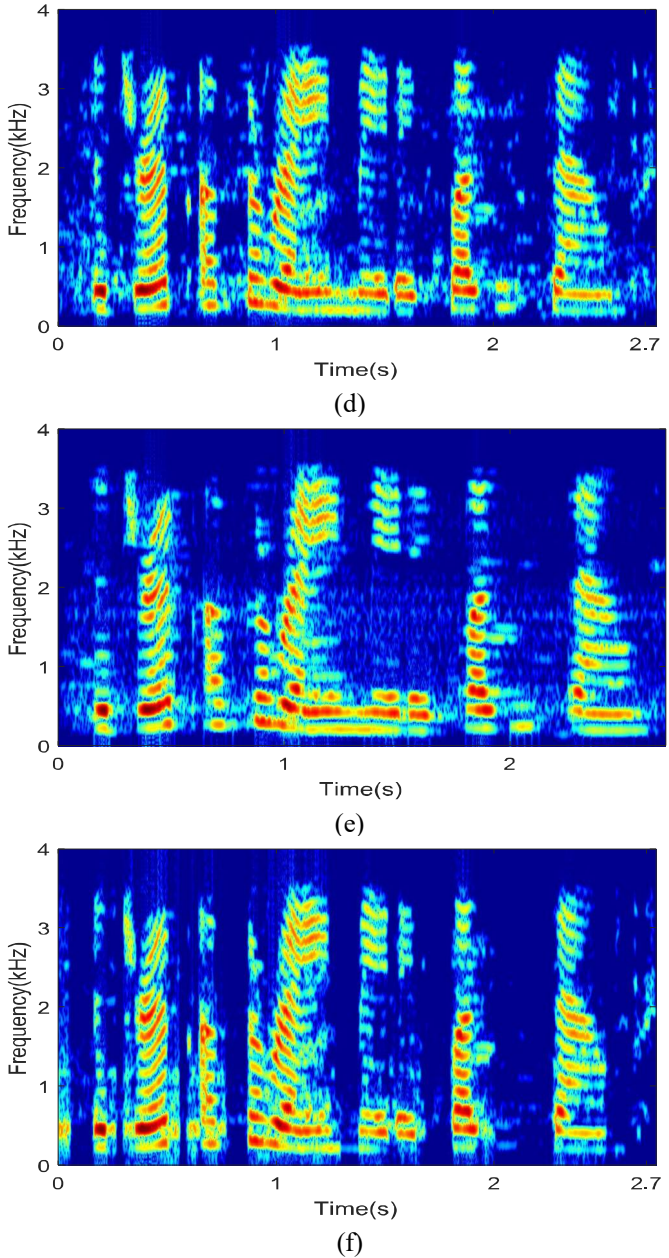

(d)

(e)

(f)

Fig. 12 (cont.)

SNR even for noisy speech corrupted with $-30$ dB street or babble noise. The simulation results show that the proposed method yields consistently better results in the sense of higher segmental SNR improvement in dB, higher overall SNR improvement and higher output PESQ values than those of the existing methods.

## REFERENCES

[1] S. Boll, "Suppression of acoustic noise in speech using spectral subtraction," *IEEE Transactions on acoustics, speech, and signal processing*, vol. 27, no. 2, pp. 113–120, 1979.

[2] K. Yamashita and T. Shimamura, "Nonstationary noise estimation using low-frequency regions for spectral subtraction," *IEEE Signal processing letters*, vol. 12, no. 6, pp. 465–468, 2005.

[3] Y. Lu and P. C. Loizou, "A geometric approach to spectral subtraction," *Speech communication*, vol. 50, no. 6, pp. 453–466, 2008.

[4] M. T. Islam, C. Shahnaz, and S. Fattah, "Speech enhancement based on a modified spectral subtraction method," in *2014 IEEE 57th International Midwest Symposium on Circuits and Systems (MWSCAS)*. IEEE, 2014, pp. 1085–1088.

[5] M. T. Islam, A. B. Hussain, K. T. Shahid, U. Saha, and C. Shahnaz, "Speech enhancement based on noise compensated magnitude spectrum," in *Informatics, Electronics Vision (ICIEV), 2014 International Conference on*. IEEE, 2014, pp. 1–5.

[6] Y. Ephraim and D. Malah, "Speech enhancement using a minimum mean-square error log-spectral amplitude estimator," *IEEE Transactions on Acoustics, Speech, and Signal Processing*, vol. 33, no. 2, pp. 443–445, 1985.

[7] Y. Hu and P. C. Loizou, "Subjective comparison and evaluation of speech enhancement algorithms," *Speech communication*, vol. 49, no. 7, pp. 588–601, 2007.

[8] Y. Ephraim and H. L. Van Trees, "A signal subspace approach for speech enhancement," *IEEE Transactions on speech and audio processing*, vol. 3, no. 4, pp. 251–266, 1995.

[9] Y. Hu and P. C. Loizou, "A generalized subspace approach for enhancing speech corrupted by colored noise," *IEEE Transactions on Speech and Audio Processing*, vol. 11, no. 4, pp. 334–341, 2003.

[10] D. L. Donoho, "De-noising by soft-thresholding," *IEEE transactions on information theory*, vol. 41, no. 3, pp. 613–627, 1995.

[11] M. Bahoura and J. Rouat, "Wavelet speech enhancement based on the teager energy operator," *IEEE Signal Process. Lett.*, vol. 8, pp. 10–12, 2001.

TABLE III: Mean scores of SIG scale for different methods in presence of street noise at 5 dB

| Listener | MBSS | PSC | SMPO | MSPP |
|---|---|---|---|---|
| 1 | 4.2 | 3.5 | 4.0 | 4.1 |
| 2 | 3.8 | 3.4 | 3.8 | 3.8 |
| 3 | 4.0 | 3.4 | 4.1 | 4.5 |
| 4 | 4.1 | 3.9 | 4.2 | 4.4 |
| 5 | 3.2 | 3.3 | 3.9 | 4.8 |
| 6 | 3.4 | 3.2 | 4.6 | 3.8 |
| 7 | 3.5 | 3.4 | 3.8 | 4.5 |
| 8 | 3.6 | 3.2 | 4.1 | 4.4 |
| 9 | 3.4 | 3.2 | 4.5 | 3.7 |
| 10 | 3.7 | 3.9 | 4.8 | 4.2 |

TABLE IV: Mean scores of BAK scale for different methods in presence of street noise at 5 dB

| Listener | MBSS | PSC | SMPO | MSPP |
|---|---|---|---|---|
| 1 | 4.2 | 4.1 | 4.5 | 5.0 |
| 2 | 4.4 | 4.2 | 4.9 | 4.7 |
| 3 | 4.1 | 4.3 | 4.4 | 4.9 |
| 4 | 4.2 | 4.5 | 4.7 | 4.7 |
| 5 | 4.2 | 4.4 | 4.8 | 4.6 |
| 6 | 4.4 | 3.7 | 4.6 | 4.8 |
| 7 | 3.2 | 3.5 | 3.9 | 4.5 |
| 8 | 4.4 | 4.2 | 4.6 | 4.5 |
| 9 | 3.9 | 3.8 | 3.8 | 4.3 |
| 10 | 4.4 | 4.1 | 4.5 | 4.4 |

TABLE V: Mean scores of OVL scale for different methods in presence of street noise at 5 dB

| Listener | MBSS | PSC | SMPO | MSPP |
|---|---|---|---|---|
| 1 | 4.2 | 2.9 | 4.0 | 4.1 |
| 2 | 3.8 | 3.8 | 3.9 | 3.8 |
| 3 | 4.6 | 3.5 | 4.0 | 4.2 |
| 4 | 4.4 | 3.5 | 4.2 | 4.4 |
| 5 | 3.5 | 3.4 | 3.8 | 4.3 |
| 6 | 4.1 | 3.3 | 3.6 | 4.8 |
| 7 | 3.2 | 3.2 | 3.8 | 4.8 |
| 8 | 4.5 | 3.8 | 3.7 | 4.5 |
| 9 | 4.4 | 3.9 | 3.9 | 4.9 |
| 10 | 4.2 | 3.4 | 3.9 | 4.9 |

TABLE VI: Mean scores of SIG scale for different methods in presence of babble noise at 5 dB

| Listener | MBSS | PSC | SMPO | MSPP |
|---|---|---|---|---|
| 1 | 4.0 | 3.6 | 4.0 | 4.2 |
| 2 | 3.9 | 3.3 | 3.9 | 3.8 |
| 3 | 4.0 | 3.9 | 4.0 | 4.3 |
| 4 | 4.2 | 3.4 | 4.2 | 4.6 |
| 5 | 3.8 | 3.2 | 3.8 | 4.2 |
| 6 | 3.6 | 2.9 | 3.6 | 3.8 |
| 7 | 3.8 | 3.8 | 3.8 | 4.7 |
| 8 | 3.6 | 3.4 | 3.6 | 4.4 |
| 9 | 3.9 | 3.5 | 3.9 | 3.8 |
| 10 | 3.8 | 3.7 | 3.8 | 3.8 |

TABLE VII: Mean scores of BAK scale for different methods in presence of babble noise at 5 dB

| Listener | MBSS | PSC | SMPO | MSPP |
|---|---|---|---|---|
| 1 | 4.5 | 4.0 | 4.5 | 4.5 |
| 2 | 4.9 | 4.3 | 4.9 | 4.5 |
| 3 | 4.4 | 4.2 | 4.4 | 4.8 |
| 4 | 4.7 | 4.4 | 4.7 | 4.9 |
| 5 | 4.8 | 4.2 | 4.8 | 4.8 |
| 6 | 4.6 | 3.9 | 4.6 | 4.6 |
| 7 | 3.9 | 3.8 | 3.9 | 4.4 |
| 8 | 4.6 | 4.4 | 4.6 | 4.8 |
| 9 | 3.9 | 3.5 | 3.9 | 4.6 |
| 10 | 4.8 | 4.7 | 4.8 | 4.7 |

TABLE VIII: Mean scores of OVL scale for different methods in presence of babble noise at 5 dB

| Listener | MBSS | PSC | SMPO | MSPP |
|---|---|---|---|---|
| 1 | 4.1 | 2.9 | 4.0 | 4.3 |
| 2 | 3.9 | 3.4 | 3.8 | 3.9 |
| 3 | 4.5 | 3.4 | 4.1 | 4.4 |
| 4 | 4.1 | 3.3 | 4.2 | 4.3 |
| 5 | 3.8 | 3.2 | 3.9 | 4.4 |
| 6 | 4.4 | 3.7 | 4.6 | 4.7 |
| 7 | 3.9 | 3.4 | 3.8 | 4.4 |
| 8 | 4.0 | 3.6 | 4.1 | 4.3 |
| 9 | 4.4 | 3.1 | 4.5 | 4.8 |
| 10 | 4.5 | 3.1 | 4.8 | 4.8 |

[12] Y. Ghanbari and M. Mollaei, "A new approach for speech enhancement based on the adaptive thresholding of the wavelet packets," *Speech Commun.*, vol. 48, pp. 927–940, 2006.
[13] M. T. Islam, C. Shahnaz, W.-P. Zhu, and M. O. Ahmad, "Speech enhancement based on student modeling of teager energy operated perceptual wavelet packet coefficients and a custom thresholding function," *IEEE/ACM Transactions on Audio, Speech, and Language Processing*, vol. 23, no. 11, pp. 1800–1811, 2015.
[14] N. Wiener, *Extrapolation, interpolation, and smoothing of stationary time series*. MIT press Cambridge, 1949, vol. 2.
[15] N. Ma, M. Bouchard, and R. A. Goubran, "Speech enhancement using a masking threshold constrained kalman filter and its heuristic implementations," *IEEE Transactions on Audio, Speech, and Language Processing*, vol. 14, no. 1, pp. 19–32, 2006.
[16] J. B. Allen and L. R. Rabiner, "A unified approach to short-time fourier analysis and synthesis," *Proceedings of the IEEE*, vol. 65, no. 11, pp. 1558–1564, 1977.
[17] R. Crochiere, "A weighted overlap-add method of short-time fourier analysis/synthesis," *IEEE Transactions on Acoustics, Speech, and Signal Processing*, vol. 28, no. 1, pp. 99–102, 1980.
[18] M. Portnoff, "Short-time fourier analysis of sampled speech," *IEEE Transactions on Acoustics, Speech, and Signal Processing*, vol. 29, no. 3, pp. 364–373, 1981.
[19] D. Griffin and J. Lim, "Signal estimation from modified short-time fourier transform," *IEEE Transactions on Acoustics, Speech, and Signal Processing*, vol. 32, no. 2, pp. 236–243, 1984.
[20] K. Wójcicki, M. Milacic, A. Stark, J. Lyons, and K. Paliwal, "Exploiting conjugate symmetry of the short-time fourier spectrum for speech enhancement," *IEEE Signal processing letters*, vol. 15, pp. 461–464, 2008.
[21] A. P. Stark, K. K. Wójcicki, J. G. Lyons, K. K. Paliwal, and K. K. Paliwal, "Noise driven short-time phase spectrum compensation procedure for speech enhancement." in *INTERSPEECH*, 2008, pp. 549–552.
[22] M. T. Islam and C. Shahnaz, "Speech enhancement based on noise-compensated phase spectrum," in *Electrical Engineering and Information & Communication Technology (ICEEICT), 2014 International Conference on*. IEEE, 2014, pp. 1–5.
[23] N. Virag, "Single channel speech enhancement based on masking properties of the human auditory system," *IEEE Transactions on speech and audio processing*, vol. 7, no. 2, pp. 126–137, 1999.
[24] S. Kamath and P. Loizou, "A multi-band spectral subtraction method for enhancing speech corrupted by colored noise," in *IEEE international conference on acoustics speech and signal processing*, vol. 4. Citeseer, 2002, pp. 4164–4164.
[25] M. Berouti, R. Schwartz, and J. Makhoul, "Enhancement of speech corrupted by acoustic noise," in *Acoustics, Speech, and Signal Processing, IEEE International Conference on ICASSP'79.*, vol. 4. IEEE, 1979, pp. 208–211.
[26] N. W. Evans, J. S. Mason, W. M. Liu, and B. Fauve, "An assessment on the fundamental limitations of spectral subtraction," in *2006 IEEE International Conference on Acoustics Speech and Signal Processing Proceedings*, vol. 1. IEEE, 2006, pp. I–I.
[27] N. B. Yoma, F. R. McInnes, and M. A. Jack, "Improving performance of spectral subtraction in speech recognition using a model for additive noise," *IEEE Transactions on speech and audio processing*, vol. 6, no. 6, pp. 579–582, 1998.
[28] I. Cohen, "Noise spectrum estimation in adverse environments: Improved minima controlled recursive averaging," *IEEE Transactions on speech and audio processing*, vol. 11, no. 5, pp. 466–475, 2003.
[29] D. O'shaughnessy, *Speech communication: human and machine*. Universities press, 1987.
[30] Y. Lu and P. C. Loizou, "Estimators of the magnitude-squared spectrum and methods for incorporating snr uncertainty," *IEEE transactions on audio, speech, and language processing*, vol. 19, no. 5, pp. 1123–1137, 2011.
[31] Y. Hu and P. C. Loizou, "Evaluation of objective quality measures for speech enhancement," *IEEE Transactions on audio, speech, and language processing*, vol. 16, no. 1, pp. 229–238, 2008.
[32] ——, "Evaluation of objective quality measures for speech enhancement," *IEEE Transactions on audio, speech, and language processing*, vol. 16, no. 1, pp. 229–238, 2008.
[33] Y. Lu and P. C. Loizou, "A geometric approach to spectral subtraction," *Speech communication*, vol. 50, no. 6, pp. 453–466, 2008.
[34] Y. Hu and P. Loizou, "Subjective comparison and evaluation of speech enhancement algorithms," *Speech Commun.*, vol. 49, pp. 588–601, 2007.
[35] ITU, "P.835 IT: subjective test methodology for evaluating speech communication systems that include noise suppression algorithms." *ITU-T Recommendation (ITU, Geneva)*, p. 835, 2003.